\documentclass[a4paper,11pt]{article}
\usepackage{aaskaiid}
\usepackage{orcidlink}

\title{Pulsar Science with the SKAO }
\ShortTitle{Pulsars with SKAO}

\author[1,2]{Bhal Chandra Joshi\orcidlink{0000-0002-0863-7781}}
\ShortName{Joshi et al.} 
\author[3]{Aris Karastergiou\orcidlink{0000-0002-1434-9786}}
\author[4]{Marta Burgay\orcidlink{0000-0002-8265-4344}}

\affiliation[1]{National Centre for Radio Astrophysics, SP Pune University Campus, Pune 411007, Maharashtra, India}
\emailAdd{bcj@ncra.tifr.res.in}
\affiliation[2]{Department of Physics, Indian Institute of Technology Roorkee, Roorkee 247667, Uttarakhand, India}
\affiliation[3]{Department of Astrophysics, University of Oxford, Denys Wilkinson Building, Keble Road, Oxford OX1 3RH, UK}
\emailAdd{aris.karastergiou@physics.ox.ac.uk}
\affiliation[4]{INAF - Osservatorio Astronomico di Cagliari, via della Scienza 5, 09047 Selargius (CA), Italy}
\emailAdd{marta.burgay@inaf.it}

\abstract{The large instantaneous sensitivity, wide frequency coverage and flexible observation modes, with large number of beams in the sky, are the main features of the SKA observatory's two telescopes, the SKA-Low and the SKA-Mid. Owing to these capabilities, the SKAO telescopes are going to be a game-changer for radio astronomy in general and pulsar astronomy in particular. Eleven chapters in this book describe their impact on different areas of pulsar science. In this overview article each chapter is briefly summarised and the inter-relationship between different pulsar science use cases are explored: new deep surveys, covering the Galactic field, globular clusters and the Galactic centre, will discover thousands of new pulsars; these will form the backbone for studies of neutron star physics and of their environments. The enhanced understanding provided by these studies will feed into the main contributions to fundamental physics from pulsar astronomy: testing relativistic gravity, studying gravitational waves in the nano-Hz regime and studying the equation of state of nuclear matter. Synergies with other science cases are also highlighted throughout this overview.}

\begin{document}
\maketitle

\section{Introduction}
The SKA Observatory (SKAO) is going to be a game-changer 
for radio astronomy in the coming decade. The construction of 
this  unique instrument, consisting of two 
telescopes in two different continents, has already commenced with 
the phase 1 providing an unprecedented seamless 
frequency coverage from 50 MHz to 15 GHz with unmatched sensitivities 
in highly flexible interferometric configurations. Therefore, the SKAO  
is expected to enable a range of diverse astronomical investigations 
with significant impact on fundamental aspects of astrophysics.

\begin{figure}[ht] 
  \centering
  \includegraphics[width=\textwidth]{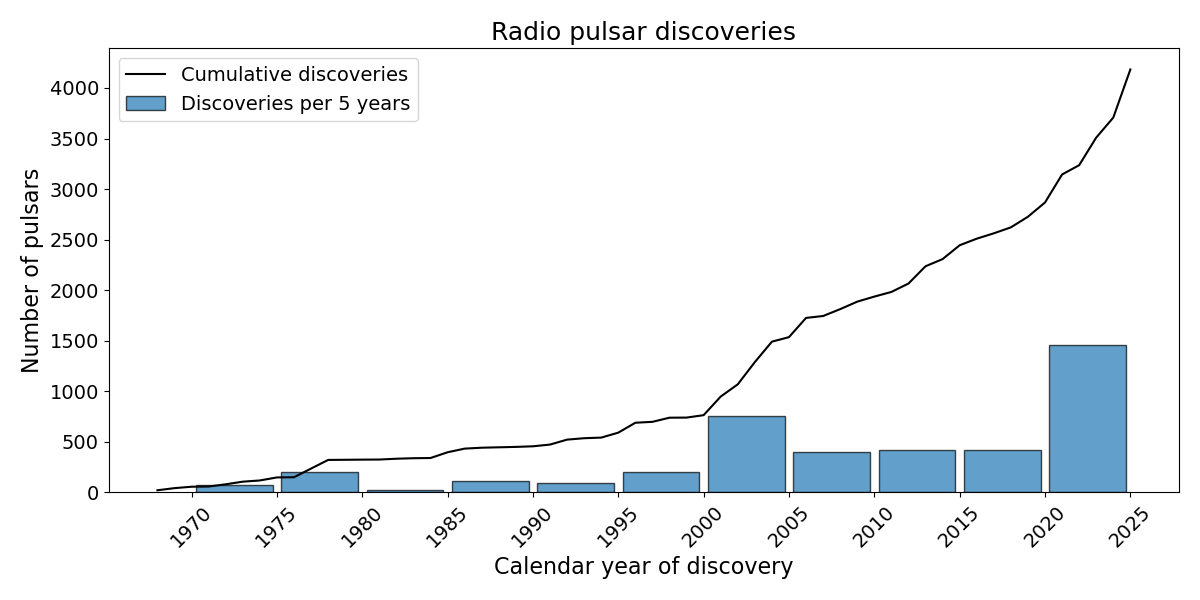}
  \caption{The discovery of pulsars has approximately occurred  
  at an exponential rate with time, matching developments in new 
  telescope hardware and software. The significant increase 
  in the last decade corresponds to commencement of operation 
  of upgraded GMRT, MeerKAT and FAST instruments. Current telescopes and future 
  assemblies of the SKAO will continue this trend.}
  \label{fig:discoveries}
\end{figure}

One such area of radio astronomy, where the SKAO is expected to make 
fundamental discoveries, is pulsar astronomy. Historically, new instruments have accelerated the pace of progress in pulsar astronomy (Figure \ref{fig:discoveries})  and the SKAO telescopes are likely to drive significant progress in this field in the next decade. The phase 1 of the rollout of the SKAO telescope, called AA4 (Array Assembly 4), is likely to double the known pulsar population in new surveys \citep{Keane01.2026.SKA,Abbate01.2026.SKA,Bagchi01.2026.SKA}. 
These new discoveries will improve our understanding of the dynamics, 
evolution  and gas content of globular clusters \citep{Bagchi01.2026.SKA} and 
the black hole at the centre of the Milky Way galaxy \citep{Abbate01.2026.SKA} 
as well as increase the samples for the different kinds of 
radio emitting neutron stars (NS) in the Galactic field, making it possible to uncover evolutionary pathways between them \citep{Levin01.2026.SKA}. 
The larger population sample will enhance our understanding of the magneto-ionic interstellar medium  \citep{Tiburzi01.2026.SKA,JunXu01.2026.SKA}, the pulsar magnetosphere \citep{Oswald01.2026.SKA} and pulsar wind nebulae  \citep{Gelfand01.2026.SKA}. 
Moreover, the discovery of exotic neutron star systems  will test 
gravity theory ever more stringently \citep{VenkatramanKrishnan01.2026.SKA}  and will probe fundamental physics at sub-atomic level \citep{AvishekBasu01.2026.SKA}. Finally, this enhanced sample is likely to make the sky portrait sharper in nano-Hertz gravitational waves impacting on our understanding of the Universe in a fundamental way \citep{Shannon01.2026.SKA}. In summary, the pulsar-related chapters in this book describe the way the upcoming SKA Observatory's telescopes address fundamental physics through the study of pulsars and gravitational waves.

\section{State of the art and expected SKAO discoveries}

\begin{figure}[ht] 
  \centering
  \includegraphics[scale=0.35]{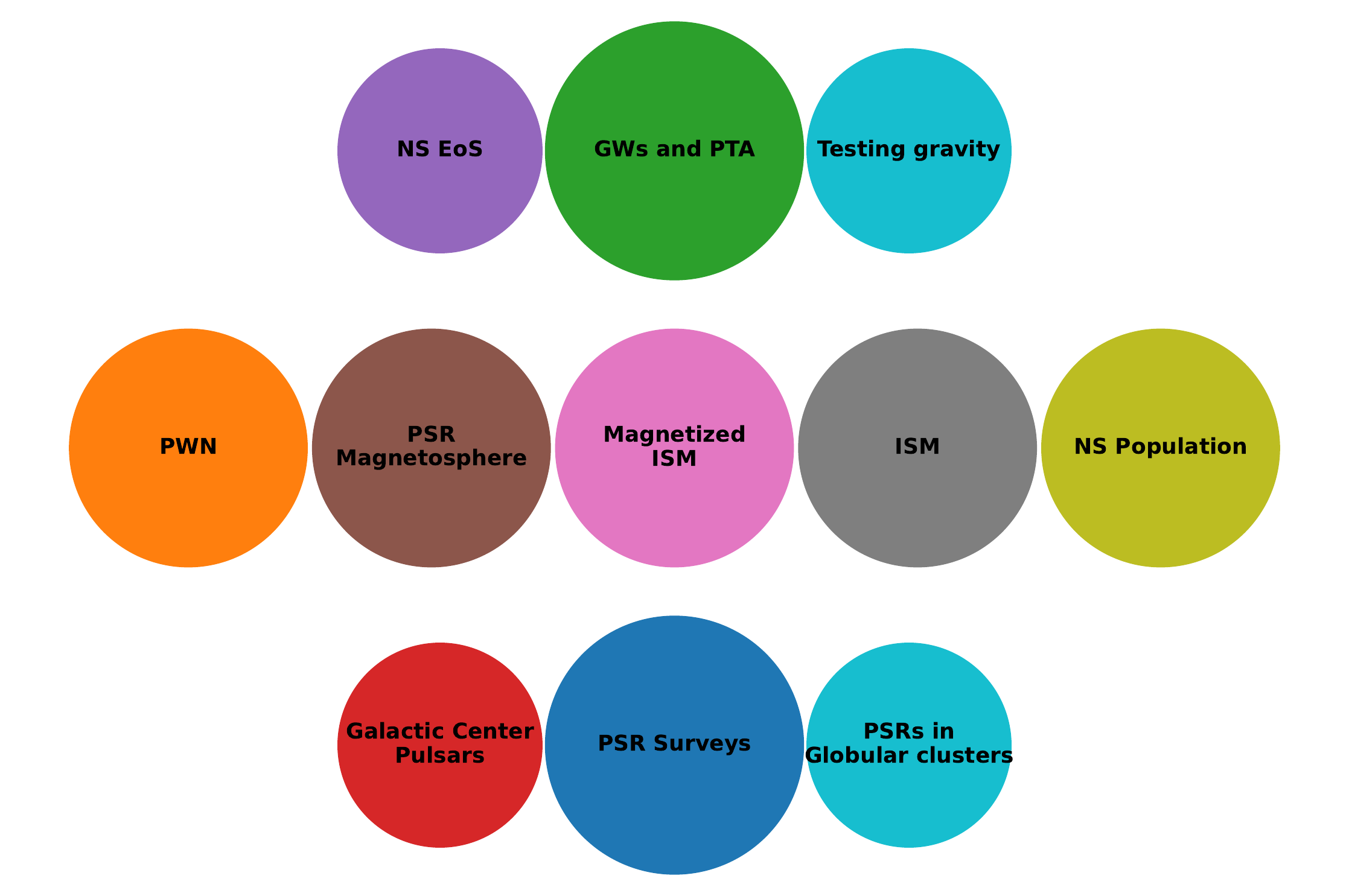}
  \caption{The inter-connected pulsar science use cases for the SKAO, described in this overview, and in the pulsar related chapters in this book, are depicted here. The surveys shown at the bottom form 
  the backbone of the study of different areas depicted in the middle row. 
  The enhanced understanding in the latter will feed into the three main 
  contributions to fundamental physics from pulsar astronomy, which are shown in the top row.}
  \label{fig:psrscience}
\end{figure}

There have been significant advances in pulsar astronomy since the pulsar science case for the SKAO was drafted two decades ago, particularly since its update a decade back \citep{2015aska.confE..36K}. Gravitational waves at high frequencies have been detected from more than 200 sources 
\citep{GWcat4} 
and the pulsar timing arrays are close to a significant detection of nano-Hertz gravitational waves 
\citep{EPTA+2023a,2023ApJ...951L...9A,2023PASA...40...49Z,2025A&A...699A.165C,2025MNRAS.536.1467M} 
opening up a messenger that the SKAO telescope can explore in earnest. 
More exotic binary systems have been discovered and mass and radius for 
several NS are measured to a good precision\footnote{\url{https://www3.mpifr-bonn.mpg.de/staff/pfreire/NS_masses.html}}, motivating investigations in 
fundamental physics using pulsars as probes 
(see e.g. \citealt{2025Ap&SS.370...74H}). 
The large instantaneous 
sensitivity and flexibility of the SKAO telescopes will have a significant 
impact on the pulsar science cases, doubling the pulsar population 
in a pulsar survey, which forms the basis for fundamental physics 
studies interconnected with other areas of pulsar astronomy, as illustrated in Figure \ref{fig:psrscience}. A brief overview of the current status of these science cases, with anticipated impact of the SKAO, is presented in the next sections summarising the details elaborated in eleven chapters presented by the pulsar Science Working Group. Synergies with science use cases of 
other science working groups is pointed out by referencing the relevant chapters in these 
sections and towards the end of the chapter.

\subsection{The SKA Pulsar Census (Keane et al.)}

Over the past fifty years, pulsar surveys have expanded the known population to over 4000 sources, revealing extraordinary diversity in spin, magnetic field, and evolutionary stage. Yet, population synthesis models predict that the Milky Way hosts more than 50,000 active pulsars. Even with modern facilities such as Murriyang (the Parkes radio telescope), FAST, and MeerKAT, much of this population---particularly faint, distant, or fast-spinning pulsars---remains undetected. A complete census of pulsars is the foundation for nearly all SKA pulsar science, underpinning studies of Galactic structure, neutron-star physics, and gravitational-wave detection.

The \textbf{SKA-Low AA*} and \textbf{SKA-Mid AA*} arrays will form a coordinated two-tier survey strategy. The SKA-Low telescope will sweep the Galactic halo and high-latitude regions at low frequencies, while the SKA-Mid telescope will target the dense Galactic plane where dispersion and scattering are strongest. Together, they are expected to detect approximately 10,000  slow pulsars and around 800 millisecond pulsars (MSPs), more than doubling the known population. This early delivery phase will already establish a statistically robust sample for pulsar timing array and population studies.

The full \textbf{AA4} configuration will increase sensitivity by 20–30\%, raising the total yield above 12,000 pulsars and providing complete coverage of the Galactic disk, bulge, and halo. The \textbf{SKA-Mid AA4} will deliver ultra-precise timing of high-dispersion and relativistic systems, while the  \textbf{SKA-Low AA4} will identify older, faint, and diffuse sources in the outer Galaxy. This comprehensive census will serve as the foundation for all subsequent SKA pulsar investigations, forming the most complete neutron-star catalogue in existence \citep{Keane01.2026.SKA}.

\subsection{Pulsars in Globular Clusters (Bagchi et al.)}

Globular clusters (GCs) are stellar systems with extreme densities, where frequent stellar encounters lead to efficient formation of millisecond pulsars through binary recycling and to the formation of exotic pairs, impossible to find in the Galactic field. Surveys with the Green Bank, and Murriyang radio telescopes and, more recently, with MeerKAT and FAST have discovered more than 350\footnote{\url{https://www3.mpifr-bonn.mpg.de/staff/pfreire/GCpsr.html}} cluster pulsars, some probing exotic regimes such as pulsar--black-hole binaries and ultradense stellar cores. Yet, modelling suggests that thousands of cluster MSPs remain hidden, limited primarily by telescope sensitivity.

The \textbf{SKA-Mid AA*} array will make immediate breakthroughs through targeted, deep searches of Galactic clusters, where only a few tied-array beams are required. Early observations are expected to more than double the number of known cluster pulsars by detecting faint and distant MSPs previously below the detection threshold. The \textbf{SKA-Low AA*} will complement these searches at lower frequencies, identifying steep-spectrum sources and diffuse emission halos around cluster cores.

Once the full \textbf{AA4} arrays are operational, the \textbf{SKA-Mid AA4} could increase the total number of detected cluster pulsars by a factor of four to five---up to 1,700 pulsars across the Galactic GC population. The \textbf{SKA-Low AA4} will simultaneously monitor multiple clusters to measure dispersion and rotation measures, mapping intra-cluster gas and magnetic fields. These data will enable precision timing of exotic binaries, exploration of cluster evolution, and tests of gravity in the densest stellar environments \citep{Bagchi01.2026.SKA}. See also \citep{Lin01.2026.SKA}.

\subsection{Galactic Centre Pulsars (Abbate et al.)}

The Galactic Centre, hosting Sagittarius~A*, is one of the most extreme gravitational environments known. Infrared monitoring of orbiting stars has confirmed the black hole’s mass and Schwarzschild precession, but pulsars---ideal clocks in curved spacetime---have yet to be found in close orbit. Despite decades of surveys with Murriyang, Effelsberg, and the Green Bank Telescope, only seven pulsars within 100~pc of the Centre are known, their detection hindered by severe radio scattering.

The \textbf{SKA-Mid AA*} configuration will overcome this barrier through its high-frequency coverage (up to 15~GHz) and sensitivity, capable of detecting heavily scattered pulsar signals near Sgr~A*. Meanwhile, the \textbf{SKA-Low AA*} will map the intervening ionized medium and measure scattering properties. 
Even during early operations, AA* is expected to uncover dozens of new pulsars in the region, expanding the sample dramatically.

At full power, the \textbf{SKA-Mid AA4} will time pulsars orbiting Sgr~A* with microsecond precision, measuring its spin and quadrupole moment---yielding direct tests of the cosmic censorship and no-hair theorems. The \textbf{SKA-Low AA4} will expand the survey area to identify 
bright transients/pulsars in the surrounding star-forming region. Together, these discoveries will transform our understanding of the Galactic Centre’s plasma physics, gravitational field, and possibly its dark-matter content \citep{Abbate01.2026.SKA}.

\subsection{The Galactic Neutron Star Population (Levin et al.)}

The past half century has revealed that neutron stars manifest in many forms---ordinary pulsars, magnetars, rotating radio transients (RRATs), intermittent pulsars, central compact objects (CCOs), and millisecond pulsars (MSPs), many of the latter in exotic binary systems, such as double neutron stars, triple system and transitional pulsars---each providing insight into extreme physics. However, biases toward bright, long-lived sources have obscured the true diversity of this population. Many faint, intermittent, or short-lived neutron stars remain undetected, preventing a unified view of their life cycles.

The \textbf{SKA-Low AA*} and the \textbf{SKA-Mid AA*} telescopes will bridge these gaps by discovering and characterizing thousands of new neutron stars across all subgroups. The \textbf{SKA-Low AA*}, with its broad field of view, will be ideal for detecting faint and steep-spectrum sources, while the  \textbf{SKA-Mid AA*} will enable high-cadence timing and polarization studies to classify and connect subpopulations. Together, these facilities will reveal transitional objects---linking, for example, high-magnetic field pulsars to magnetars or RRATs to ordinary pulsars---and provide the first coherent picture of neutron-star evolution.

In the \textbf{AA4} era, \textbf{SKA-Mid} will conduct continuous timing and polarization monitoring of thousands of sources, measuring glitches, emission variability, and magnetospheric changes, while \textbf{SKA-Low} will expand discovery surveys into the Galactic halo. These combined efforts will yield the first statistically complete neutron-star census, tracing evolutionary pathways and unifying disparate classes within a single population framework \citep{Levin01.2026.SKA}.

\subsection{Pulsar Magnetospheres (Oswald et al.)}

Advances in radio and high-energy astrophysics have revealed that pulsar magnetospheres exhibit extraordinary complexity, including non-dipolar fields, plasma instabilities, and emission variability. Observations with FAST, MeerKAT, and NICER have deepened our understanding, but left the core questions---how and where coherent radio emission arises---unanswered.

The \textbf{SKA-Mid AA*} array, with its wide bandwidth and unmatched frequency coverage, particularly using sub-arrays for concurrent multi-band observations together with its  large sensitivity, will conduct high-cadence monitoring of polarization and single-pulse behaviour in hundreds of pulsars. This will constrain magnetic geometry and emission altitude, while the \textbf{SKA-Low AA*} will probe coherent plasma processes near the stellar surface. Together, the SKA-Low and SKA-Mid arrays will bridge observational and theoretical gaps in pulsar emission physics.

In the \textbf{AA4} configuration, the \textbf{SKA-Mid} will monitor thousands of pulsars at sub-millisecond resolution, uncovering population-wide variability patterns, while the \textbf{SKA-Low} will capture weak, intermittent, or low-frequency emission signatures. These results will refine models of magnetospheric structure, test theories of particle acceleration, and link observed radio emission directly to fundamental plasma dynamics \citep{Oswald01.2026.SKA}.

\subsection{Galactic Plasma Studies (Tiburzi et al.)}

Pulsars have long served as precise probes of the Galaxy’s ionized plasma. Observations with LOFAR, GMRT, and MeerKAT have traced electron-density fluctuations and revealed turbulence consistent with a Kolmogorov spectrum, while long-term dispersion measure (DM) monitoring has exposed changes due to solar wind and interstellar structures. Yet, these measurements are limited in frequency coverage and spatial sampling.

The \textbf{SKA-Low AA*} array will achieve unmatched precision in dispersion and scattering studies, tracking DM variations across hundreds of channels and enabling detailed mapping of the turbulent interstellar medium (ISM). In parallel, the \textbf{SKA-Mid AA*} will provide high-frequency coverage to separate frequency-dependent dispersion from refraction, allowing a full multi-scale characterization of Galactic plasma and solar-wind dynamics.

Once \textbf{AA4} arrays are in place, the \textbf{SKA-Low} will deliver a high-resolution 3D model of Galactic electron density, while the \textbf{SKA-Mid} will furnish ultra-stable timing data critical to Pulsar Timing Arrays. A complimentary dimension will be added by potential space VLBI observations with the SKAO telescopes combining ultra-precise astrometry parallaxes of pulsars with scattering measurements using pulsars \citep{Kovalev01.2026.SKA}. Together, these observations will refine models of plasma turbulence, correct for ISM effects in gravitational-wave searches, and provide the most comprehensive view to date of the Milky Way’s magneto-ionic medium \citep{Tiburzi01.2026.SKA}.

\subsection{Galactic Magnetic Fields (Xu et al.)}

The study of Galactic magnetism has progressed from optical polarization measurements to radio Faraday rotation mapping using pulsars and extragalactic sources. Large scale experiments with Murriyang, FAST and MeerKAT have revealed large-scale field reversals and magnetized halo structures, yet data remain confined to the local half of the disk. The three-dimensional structure and global symmetry of the Milky Way’s magnetic field remain incompletely characterized.

The \textbf{SKA-Mid AA*} and \textbf{SKA-Low AA*} arrays will transform this picture. Early-phase (AA*) observations will triple the number of pulsars with measured rotation measures, providing thousands of new magnetic-field sight-lines across the disk and halo. The \textbf{SKA-Mid AA*} will target the inner spiral arms, while the \textbf{SKA-Low AA*} will probe the diffuse magnetism of the Galactic halo and high-latitude sky, extending magnetic mapping to kiloparsec scales.

With the full \textbf{AA4} arrays, the SKAO telescopes will deliver a much improved 3D magnetic-field map of the Milky Way. The \textbf{SKA-Mid AA4} will measure field reversals and small-scale turbulence, and the \textbf{SKA-Low AA4} will chart the extended halo field. The complimentary observations from other
proposed  polarization surveys will reveal the most complete picture of the Galactic magnetic 
field \citep[See also][]{Pathak01.2026.SKA,Ma01.2026.SKA,Tahani01.2026.SKA}.

\subsection{Pulsar Wind Nebulae (Gelfand et al.)}

Pulsar wind nebulae (PWNe) illustrate how neutron stars transfer rotational energy to their surroundings, producing non-thermal emission across the electromagnetic spectrum. Observations with Chandra, H.E.S.S., and MeerKAT have revealed spectacular structures such as jets, tori, and bow shocks, but most PWNe remain unresolved, and their particle acceleration mechanisms are not yet fully understood.

The \textbf{SKA-Mid AA*} array will provide deep, wide-band imaging and polarization mapping of PWNe, tracing their magnetic topology and energy distribution. \textbf{SKA-Low AA*} will detect diffuse, low-frequency synchrotron emission from older or larger nebulae, capturing relic populations of relativistic electrons. Joint SKA-Low and SKA-Mid observations will reveal how magnetic structure and particle spectra evolve over time.

At full capacity, \textbf{SKA-Mid AA4} will resolve fine-scale features---filaments, shocks, and jets---at sub-arcsecond resolution, while \textbf{SKA-Low AA4} will quantify large-scale halos and measure depolarization to infer field geometry. VLBI with the SKA  will reveal emergence of nascent PWNs \citep{TaoAn02.2026.SKA}. Together, these data will clarify how pulsar winds accelerate particles to PeV energies, their role in cosmic-ray production, and their feedback on the interstellar medium \citep{Gelfand01.2026.SKA}.

\subsection{Testing Gravity with Binary Pulsars (Venkatraman Krishnan et al.)}

Binary pulsars are unique laboratories for testing general relativity (GR) and alternative theories of gravity in the strong-field regime. From the orbital decay of the Hulse--Taylor pulsar to the double pulsar PSR~J0737--3039A/B, such systems have validated gravitational-wave emission and post-Keplerian dynamics to extraordinary precision. However, only a limited number of relativistic binaries are known, and pulsar--black-hole systems---crucial for probing spacetime curvature and the no-hair theorem---have yet to be discovered.

The \textbf{SKA-Mid AA*} array will revolutionize this field by increasing timing precision by an order of magnitude and detecting many new relativistic binaries. Early AA* observations will measure minute effects such as Shapiro delay, gravitational redshift, and frame dragging in known systems. In parallel, the \textbf{SKA-Low AA*} will survey for long-period and eccentric binaries, populating the full parameter space of relativistic systems for gravity tests. The timing precision will also be significantly improved incorporating systematics due to propagation effects in the interstellar medium using the lower frequencies provided by both the \textbf{SKA-Mid} and \textbf{SKA-Low} telescopes. 

The large instantaneous sensitivity of \textbf{AA*} and \textbf{AA4} telescopes will lead to discovery of systems with orbital period shorter than 2 hours and higher eccentricities, probing larger curvature parameter space in relatively shorter observing campaigns. They will also allow for more than a factor of two improvement in pulse times of arrival  uncertainty and an order of magnitude improvement in the precision of 
Post-Keplerian (PK) parameters, even with integration time as small as 30 s. With the \textbf{AA4} upgrade, \textbf{SKA-Mid} will achieve tens-of-nanosecond timing accuracy, allowing detection of higher-order post-Newtonian corrections and direct verification of GR’s predictions for black-hole multipole moments. \textbf{SKA-Low AA4} will extend the search to the Galactic periphery, identifying rare systems that complement timing-array experiments.  Combining astrometric orbital monitoring of the pulsar companion through VLBI with precesion pulsar timing, the astrometric distances can be determined up to 6 kpc for systems ranging from millisecond pulsar binaries to pulsar--black-hole systems \citep{Lin01.2026.SKA}. Combined, these datasets will establish SKA as the premier facility for experimental gravitation, exploring the limits of Einstein’s theory with unmatched precision \citep{VenkatramanKrishnan01.2026.SKA}.

\subsection{Probing Neutron Star Interiors and Dense Matter (Basu et al.)}

Observations over the last decade have established neutron stars as laboratories for ultra-dense matter physics at relatively low temperatures and large proton-neutron asymmetries. Pulse modelling with NICER and gravitational-wave detections, along with radio timing observations of binary neutron stars, have constrained masses, radii and moment of inertia. The timing studies of young pulsars have revealed superfluidity, crustal entrainment  and internal coupling. Nevertheless, the microphysics of matter above nuclear saturation density---and the transition between hadronic and exotic phases---remains uncertain.

The \textbf{SKA-Mid AA*} system will immediately improve mass and spin measurements by timing massive and rapidly rotating pulsars at sub-microsecond precision. The \textbf{SKA-Low AA*} will track glitches and precession events along with variation in propagation delays in the interstellar medium, providing higher precision measurements of these quantities. It will also allow us to better characterize the timing noise of these young pulsars, thereby revealing superfluid dynamics within neutron-star interiors. Early AA* results will narrow the range of viable equations of state (EoS) and test theoretical models of nuclear interactions and phase transitions.

At full \textbf{AA4} sensitivity, the \textbf{SKA-Mid} will enable high precision moment-of-inertia measurements in relativistic binaries and explore frame-dragging effects linked to internal structure. The \textbf{SKA-Low AA4}, operating at lower frequencies, will characterize glitch recovery and long-term timing noise, probing superfluid vortices and crust elasticity. An observing program resolving rotational variations on timescales from seconds to years with dynamic high cadence scheduling 
near a glitch for a large sample of pulsars with \textbf{AA4} would be fruitful and commensality with other observations would be recommended. The mass priors and geometric/polar cap priors as well as precision radio ephmerides provided by \textbf{AA4} telescopes will benefit multi-messenger observations with upcoming facilities, such as eXTP, New Athena, LISA, Cosmic Explorer and Einstein Telescope to provide tight constraints on the EoS. Therefore, the SKAO telescopes will provide definitive observational constraints on the EoS and the behaviour of matter at supra-nuclear densities \citep{AvishekBasu01.2026.SKA}.

\subsection{The SKAO Pulsar Timing Array (Shannon et al.)}

The spatial correlation introduced by gravitational waves (GWs) in the precision timing of an ensemble of millisecond pulsars, or a pulsar timing array (PTA), provides a complementary view of the Universe to that seen in decahertz GWs by terrestrial detectors, such as advanced LIGO and Virgo. While a compelling evidence was arising in decade long data released by the existing PTA experiments three years back, a significant detection -- and indeed understanding the origins of these GWs -- is yet to happen. In future, PTAs are likely to be able to discriminate between an astrophysical or cosmological origin of the stochastic gravitational background, validate and refine hierarchical galaxy evolution models, confirm or rule out anisotropy in the GW  background, set stringent limits on the dipolar GW radiation and discriminate between alternative gravity theories with a rich dividend for both astrophysics and fundamental physics. This requires extending the current time baseline in the coming decades, as well as increasing the sensitivity of a PTA experiment. This was the prime aim of pulsar science case for the SKAO telescopes, bringing together gains from  sensitive pulsar surveys and advances in pulsar EoS, interstellar medium and pulsar emission physics. 

The \textbf{SKA-Mid} and \textbf{SKA-Low} telescopes will play a pivotal role in this respect. The \textbf{SKA-Mid AA*} array will increase the MSP ensemble by both extending the time baselines for the currently monitored MSPs at the MeerKAT, Parkes pulsar timing array, North American Nano-Hertz Gravitational wave Observatory, European pulsar timing array, the Indo-Japanese pulsar timing array and the Chinese pulsar timing array, and by including hitherto inaccessible weaker MSPs. The sample will be augmented by new discoveries in the pulsar census \citep{Keane01.2026.SKA} with the \textbf{SKA-Mid AA*} and \textbf{SKA-Low AA*}. The SKAO PTA will also benefit by better modelling of chromatic effects using wide frequency coverage provided by both the \textbf{SKA-Mid} and \textbf{SKA-Low} telescopes; the latter, in particular, will provide very precise time-dependent DM measurements. The high istantaneous sensitivity of both telescopes, moreover, will allow for a better characterization of pulse jitter.

With the full \textbf{AA4}, the \textbf{SKA-Mid} and \textbf{SKA-Low} telescopes will further provide  sub-$\mu$s precision times-of-arrival for more than double the number of currently monitored pulsars with a modest use of SKAO telescope time, largely due to a factor of four times the MeerKAT sensitivity. The lower observing frequency band of the \textbf{SKA-Low AA4} array will better characterize the chromatic delays, pulse-shape evolution, and DM variability \citep{Tiburzi01.2026.SKA} thereby accounting for the noise budget more accurately \citep{Shannon01.2026.SKA}.

\section*{Synergies with other science applications of the SKAO telescopes}

The pulsar science cases outlined above either benefit or are complemented by several other science applications described in other chapters of this book.  These range from complementing other polarization surveys \citep{JunXu01.2026.SKA,Ma01.2026.SKA,Tahani01.2026.SKA},  determining astrometric masses with precision astrometry using VLBI with pulsar timing \citep{VenkatramanKrishnan01.2026.SKA,Lin01.2026.SKA}, search for radio signals from axion like particle \citep{Regis01.2026.SKA}, precision pulsar parallax measurements using multi-view techniques and differential precision astrometry \citep{Timmerman01.2026.SKA,Kobayashi01.2026.SKA,MingHuiXu01.2026.SKA,Rioja01.2026.SKA} to better localization of continuous GW sources using accurate distance measurements with VLBI measurements \citep{Takahashi01.2026.SKA}. Microhertz GWs can be sensed by precision pulsar timing measurements of resonantly enhanced secularly accumulating imprint on binary orbital frequencies of about 5 systems with many more likely to be discovered in SKAO pulsar surveys \citep{Jenkins01.2026.SKA}. Identification of continuous gravitational wave sources is likely to be enabled by tracing the sub-pc scale orbital motion using earth or space based VLBI with the SKAO telescopes \citep{DAmato01.2026.SKA}. The SKAO telescope also provide excellent instruments for multi-messenger and multi-band studies of such GW sources using pulsar timing arrays, continuum imaging and transient studies \citep{Capelo01.2026.SKA}. Pulsar timing array data can be also used in conjunction with astrometric surveys to provide tests of the possible cosmological origin of a GW background through the detection of the kinematic dipole induced by the Solar System's motion \citep{Cruz01.2026.SKA,Ragavendra01.2026.SKA} and tests of General theory of relativity \citep{Besancon01.2026.SKA} thereby probing the foundations of $\Lambda-$CDM models \citep{Camera01.2026.SKA}; and to probe first-order cosmological phase transitions  \citep{Pasechnik01.2026.SKA}

\section*{Summary}

Together, these studies define the pulsar and neutron-star science programme for the SKA Observatory. The \textbf{AA*} arrays will provide an immediate leap in discovery power and precision, while the full \textbf{AA4} configurations of \textbf{SKA-Low} and \textbf{SKA-Mid} will deliver transformative, population-level insights into neutron-star physics, plasma astrophysics, Galactic structure, and gravitation.


\section*{Acknowledgements}

BCJ acknowledges the support from Raja Ramanna Chair fellowship of the Department of Atomic Energy, Government of India (RRC - Track I Grant 3/3401 Atomic Energy Research 00 004 Research and Development 27 02 31 1002//2/2023/RRC/R\&D-II/13886 and 1002/2/2023/RRC/R\&D-II/14369). 
MB acknowledges resources from the research grant “iPeska” (PI: Possenti), funded under the INAF national call Prin-SKA/CTA approved with the Presidential Decree 70/2016.

\bibliographystyle{abbrvnat-maxbibnames4}
\bibliography{newest_chapter} 

\end{document}